\documentclass[aps,prb,reprint,superscriptaddress,raggedbottom,floatfix,longbibliography]{revtex4-2}

\usepackage[utf8]{inputenc}
\usepackage{amsmath}
\usepackage{amssymb}
\usepackage{comment}
\usepackage[pdftex]{graphics}
\usepackage{epsfig}
\usepackage{xspace}
\usepackage[usenames,dvipsnames]{xcolor}
\usepackage{MnSymbol}
\usepackage{pifont}
\usepackage[geometry]{ifsym} 
\usepackage{float}
\usepackage{etoolbox}
\newcommand{\beq}{\begin{equation}}
\newcommand{\eeq}{\end{equation}}
\newcommand{\kVcm}{\ensuremath{~\mathrm{kV}\mathrm{cm}^{-1}}} 
\newcommand{\MVcm}{\ensuremath{\mathrm{MV}\,\mathrm{cm}^{-1}}}
\newcommand{\mVx}[1]{\ensuremath{~\mathrm{m}^{#1}\,\mathrm{V}^{-#1}}}
\newcommand{\mVctn}{\ensuremath{~\mathrm{m}^5\,
\mathrm{V}^{-2}}} 
\newcommand{\pcmc}{\ensuremath{~\mathrm{cm}^{-3}}}
\newcommand{\pcms}{\ensuremath{~\mathrm{cm}^{-2}}}
\newcommand{\fwhm}{\mathrm{FWHM}}
\newcommand{\chin}[1]{\ensuremath{\chi^{(#1)}}}
\newcommand{\banh}{BNP}
\newcommand{\edsr}{EDS}
\newcommand{\Eopph}{\ensuremath{\mathcal{E}_{op}}}
\newcommand{\mum}{\ensuremath{~\mu\mathrm{m}}}
\newcommand{\muJ}{\ensuremath{~\mu\mathrm{J}}}
\newcommand{\meV}{\ensuremath{~\mathrm{meV}}}
\newcommand{\thz}{\ensuremath{~\mathrm{THz}}}
\newcommand{\hho}{\ensuremath{\mathrm{hh}}}
\newcommand{\lho}{\ensuremath{\mathrm{lh}}}
\newcommand{\soho}{\ensuremath{\mathrm{so}}}
\newcommand{\accB}[1]{\ensuremath{\dot{v}_{#1}^{\mathrm{(B)}}}}
\newcommand{\accIB}[1]{\ensuremath{\dot{v}_{#1}^{\mathrm{(IB)}}}}

\usepackage{tabularx}
\usepackage[normalem]{ulem}
\usepackage{ulem,xpatch}

\newcommand{\ffm}{Physikalisches Institut, J. W. Goethe-Universit\"at, Max-von-Laue-Strasse 1, 60438 Frankfurt am Main, Germany} 
\newcommand{\hzdr}{Helmholtz-Zentrum Dresden-Rossendorf, Bautzner Landstrasse 400, 01328  Dresden, Germany}
\newcommand{\rwth}{Institut für Theoretische Elektrotechnik, RWTH Aachen, 52062 Aachen, Germany}
\newcommand{\dlr}{Institute of Optical Sensor Systems, German Aerospace Center (DLR), 12489 Berlin, Germany}
\newcommand{\hub}{Institut für Physik, Humboldt-Universität zu Berlin, 12489 Berlin, Germany}
\newcommand{\ikz}{Leibniz-Institut für Kristallzüchtung (IKZ), 12489 Berlin, Germany}

\begin{document}


\title{Higher-harmonic generation in boron-doped silicon from band carriers and bound-dopant photoionization}


\author{Fanqi Meng}
\thanks{f.meng@physik.uni-frankfurt.de}
\author{Frederik Walla}%
\affiliation{\ffm}
\author{Sergey Kovalev}
\author{Jan-Christoph Deinert}
\author{Igor Ilyakov}
\author{Min Chen}
\author{Alexey Ponomaryov}
\affiliation{\hzdr}
\author{Sergey G. Pavlov}
\affiliation{\dlr}
\author{Heinz-Wilhelm Hübers}
\affiliation{\dlr}
\affiliation{\hub}
\author{Nikolay V. Abrosimov}
\affiliation{\ikz}
\author{Christoph Jungemann}
\affiliation{\rwth}
\thanks{cj@ithe.rwth-aachen.de}
\author{Hartmut G. Roskos}
\thanks{roskos@physik.uni-frankfurt.de}
\affiliation{\ffm}
\author{Mark D. Thomson}
\thanks{thomson@physik.uni-frankfurt.de}
\affiliation{\ffm}


\date{\today}

\begin{abstract}
We investigate ultrafast harmonic generation (HG) in Si:B, driven by 
intense pump pulses with fields reaching $\sim$100\kVcm{} and a carrier frequency of 300~GHz, at 4~K and 300~K, both experimentally and theoretically.
We report several novel findings concerning the nonlinear charge carrier dynamics in intense sub-THz fields. (i)  
Harmonics of order up to $n=9$ are observed at room temperature, while at low temperature we can resolve harmonics reaching even $n=13$. The susceptibility per charge carrier at moderate field strength is as high as for charge carriers in graphene, considered to be one of the materials with the strongest sub-THz nonlinear response. %
(ii) For $T=300$~K, where the charge carriers bound to acceptors are fully thermally ionized into the valence subbands, the susceptibility values decrease with increasing field strength. Simulations incorporating multi-valence-band Monte-Carlo and finite-difference-time-domain (FDTD) propagation show that here, the HG process becomes increasingly dominated by energy-dependent scattering rates over the contribution from band non-parabolicity, due to the onset of optical-phonon emission, which ultimately leads to the saturation at high fields. 
(iii) At $T=4$~K, where the majority of charges are bound to acceptors, we observe a drastic rise of the HG yields for internal pump fields of $\sim$30\kVcm, as one reaches the threshold for tunnel ionization.  We disentangle the HG contributions in this case into contributions from the initial ``generational"- and subsequent band-nonlinearities, and show that scattering seriously degrades any coherent recollision during the subsequent oscillation of the holes.
\end{abstract}  

\maketitle

\section{Introduction}

The future of semiconductor electronic devices relies on continuing to test and refine the physical description of carrier dynamics for increasingly shorter time scales and higher electric field strengths (i.e., sub-picosecond times and approaching the \MVcm-range).
Experiments with ultrashort, high-field pulses, in particular observing nonlinear higher order harmonic generation (HG) \cite{Meng22,Hafe18, Deinert2020,Kovalev2020,Dessmann2021}, provide a sensitive probe into the carrier response, as these nonlinear currents are dictated by the precise carrier distribution and scattering processes, band structure, interactions between band carriers and dopant/impurity ions, and, at low-temperatures (where charges are bound to acceptors), field-driven ionization.

In the journey of HG from free-carriers in semiconductors, pioneering studies using mid- \cite{Pate66} and far-infrared \cite{Maye86,Urba95} pulses from molecular lasers were generally restricted to $\sim$nanosecond pulses. Although the estimated peak electric field reached over $\sim$100\kVcm, no harmonics higher than third order were reported, and the sub-picosecond dynamics were not resolved.
Nevertheless, they fueled the development of theoretical descriptions, including the role of band non-parobolicity and carrier relaxation \cite{Char1969,Brazis1998}. 
The availability of high-field (sub-)picosecond THz pulses (e.g. from free-electron lasers or femtosecond-amplifier-laser sources) allows one to finally reach the non-perturbative field regime on the native time scales for the carrier dynamics \cite{Meng22,Dessmann2021}.
Besides fundamental interest, the HG process can be considered for practical use for frequency conversion or frequency comb generation, and the efficiency can be enhanced, e.g., using resonant cavities \cite{Meng2020}.

In this publication, we investigate HG in weakly p-doped bulk silicon in a focused 0.3-THz radiation field reaching maximum incident peak fields of $\sim$100\kVcm{} (all field values here and in the following specify the field strength in vacuum, as the field strength within the specimens varies locally due to the formation of standing waves). These fields are strong enough to induce tunnel ionization of the impurity atoms at low temperature and to drive charge carriers out in $k$-space to sufficiently high kinetic energies to reach the threshold for scattering by optical phonon emission (energy of the optical phonons at the $\Gamma$ point: $\Eopph=63.3$~meV -- the TO and LO phonons being degenerate there \cite{Dargys1996}).

\begin{figure}[!t]
	\includegraphics[keepaspectratio,width=0.35\textwidth]{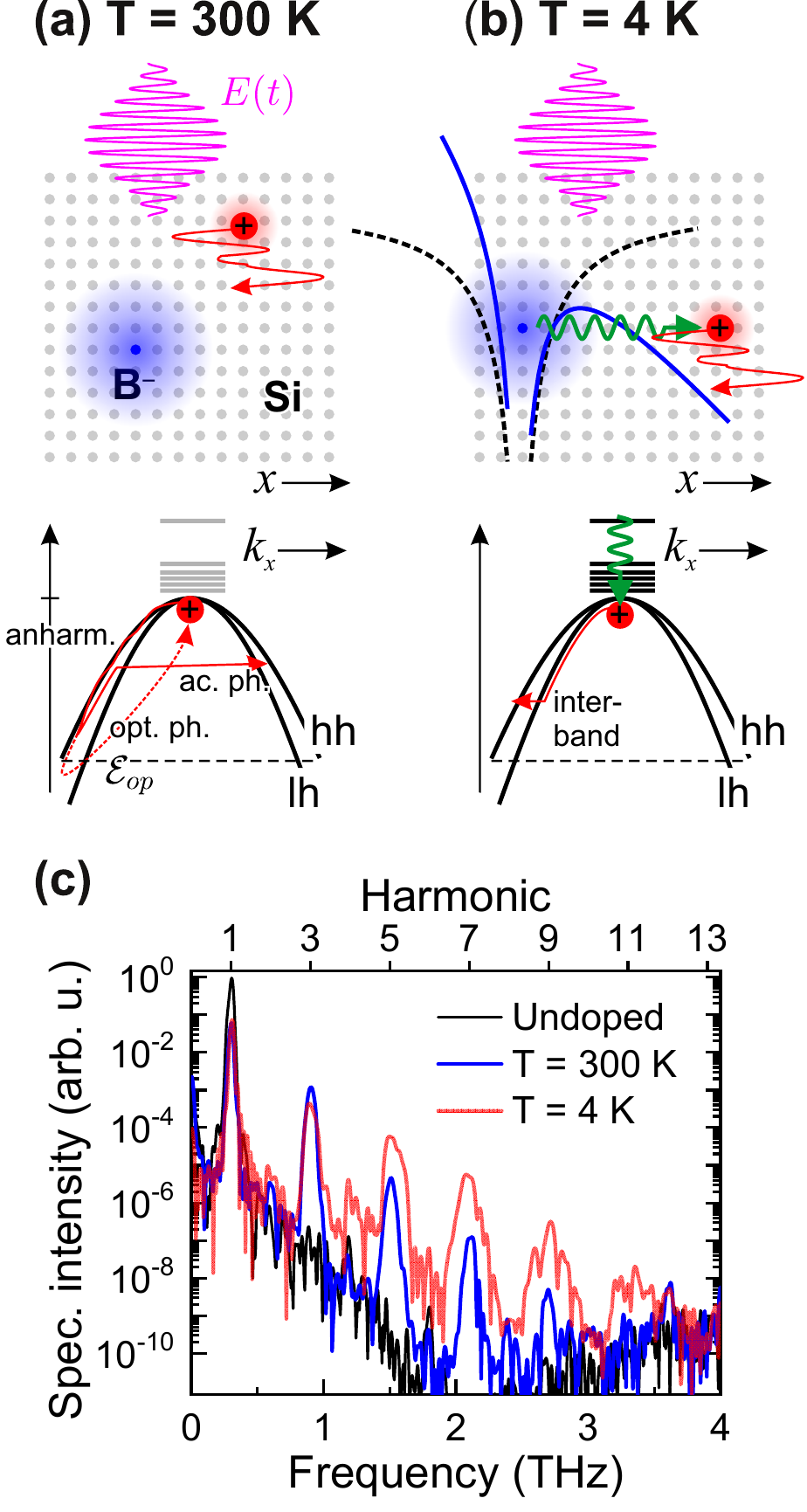}
	\caption{\label{fig:overview} Overview of strong-field harmonic generation in Si:B at (a) $T=300$~K -- band motion with thermally ionized holes including both band non-parabolicity and energy-dependent scattering; (b) $T=4$~K, whereby light holes are first generated by tunnel ionization of acceptors, followed by significant lh$\rightarrow$hh scattering (amplitude of hole trajectories not to scale).
	(c) Example of experimental emission intensity spectra for both $T=300$~K (peak external incident pump field $E_0=115\kVcm$) and $T=4$~K ($E_0=81\kVcm$) (linear transmitted fundamental spectrum with undoped sample included for comparison).
	}
\end{figure}


\section{Experimental}

The experiments were conducted at the HZDR (Dresden) by employing the TELBE superradiant undulator source, where we used linearly polarized THz pulses tuned to a carrier frequency of $\nu_1=$~300~GHz, with a pulse duration of $T_1=$~14.1~ps (FWHM) and pulse energies up to 1\muJ{} at a 50-kHz repetition rate.
We performed two series of experiments with two Si:B samples: (i) at $T=300$~K ($N_d=5.75\cdot10^{15}\pcmc$, thickness $L=900$~µm) where the dopant acceptors are fully ionized and one measures the nonlinear response of a constant population of band holes, and (ii) at $T=4$~K ($N_d=5.0\cdot10^{16}\pcmc$, thickness $L=272$~µm) where essentially all holes are bound to their parent ions and are photoionized into the valence bands during the pump pulse.
Preceding the sample, we employed a low-pass filter to suppress residual harmonics from the source, followed by two polarizers \cite{Hafe18} to allow continuous variation of the strength of the pump field in the sample. 
Following the sample, a calibrated high-pass filter \cite{Meng22} was used to reduce the amplitude of the fundamental to allow a more balanced signal level for the overtones. The emitted field from the sample was reimaged into an electro-optic (EO) crystal for coherent time-domain detection.
Incident and detected on-axis temporal field strengths were determined by calibrating the reference signal with the fluence determined from an additional measurement of the pump beam profile.
Details of the samples and determination of doping concentration, field calibration, and spectral correction for the filters and EO-response are given in the Supplementary.

\section{Results}
\subsection{Overview of the findings}\label{sec:overview}
Figure~\ref{fig:overview} shows conceptual aspects of the experiments performed at 300~K (Fig.~\ref{fig:overview}(a)) and 4~K (Fig.~\ref{fig:overview}(b)) together with spectra recorded at the two temperatures at the maximal field strength available in the experiments (Fig.~\ref{fig:overview}(c)). At room temperature, the impurities are ionized and one observes the signatures of the nonlinear response of the charge carriers accelerating and scattering in the incident THz field (linearly polarized in $x$-direction, propagating in $z$-direction). As indicated in the lower panel of Fig.~\ref{fig:overview}(a), the nonlinear response is determined by the non-parabolicity 
of the heavy- and light-hole valence bands and by the \textit{relaxational nonlinearity} associated with the absorption and emission of acoustic and optical phonons, which gives rise to HG due to the energy dependence of the scattering rates. Charged impurity scattering turns out not to  play a strong role for the nonlinear response.
At low $T$, most of the impurity atoms are initially in the neutral state. When they are exposed to a radiation field with rising field amplitude, one observes an increase of the nonlinear susceptibility $\chi^{(n)}$, which originates from the growth of the number of free holes by impact ionization and -- setting in at fields of $\sim$30\kVcm{} -- by additional tunnel ionization, as indicated in Fig.~\ref{fig:overview}(b). The tunneling process mainly populates the light-hole states from which scattering rapidly redistributes the holes in the valence band. 
We show that near the ionization threshold, a significant HG contribution is due to these ``generational" nonlinear currents, while for higher pump fields the subsequent band nonlinearities dominate.

\subsection{Room-temperature harmonic generation}\label{sec:hg300K}

\begin{figure}[!t]
	\includegraphics[keepaspectratio,width=0.5\textwidth]{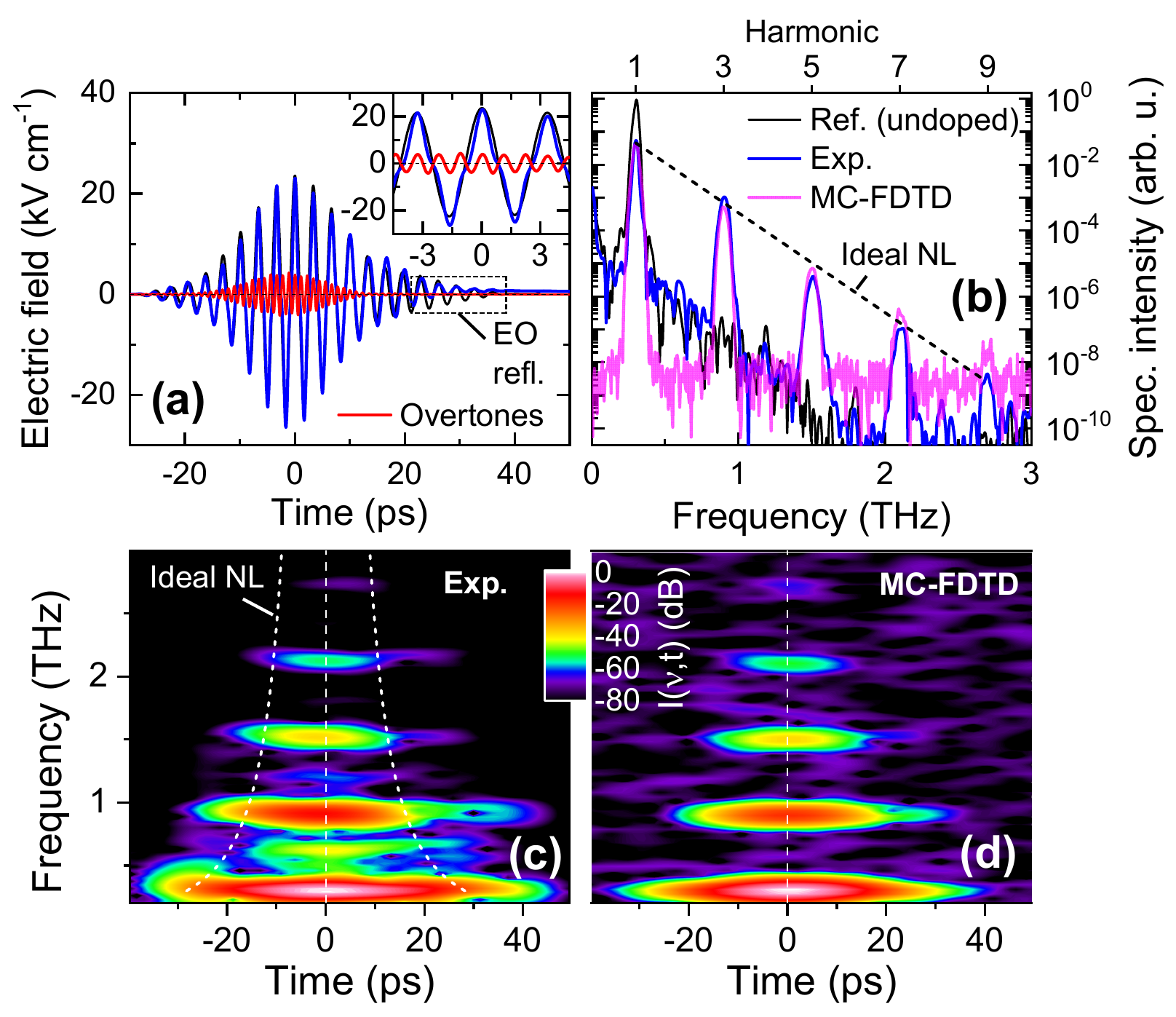}
	\caption{\label{fig:fields} (a) Experimental transmitted temporal field for a Si:B sample ($N_d=5.75\cdot10^{15}\pcmc$, $T=300$~K) for an incident pump peak field of $E_0=115\kVcm$ (blue curve), as well 
	as the field of the overtones without fundamental (red) and 
	scaled reference data for an undoped sample (black).  Inset shows detail around the pulse peak.
	(b) Corresponding intensity spectra, as well as spectrum from MC-FDTD simulations with $E_0=100\kVcm$ (see text for discussion of pump field scaling between theory and experiment). (c) Experimental and (d) MC-FDTD spectrograms (common color scale as shown). Included in (b) and (c) are dashed curves for the relative intensity and duration, respectively, for ideal nonlinear HG (for the latter, assuming a Gaussian pulse profile, such that $T_n\sim T_1/\sqrt{n}$, curves shown for $T_n=\pm 2T_\fwhm$).
	}
\end{figure}

We begin with the detailed description of the results measured at $T=300$~K, to study the nonlinear response of a constant density of thermally ionized band holes.
Examples of the fields measured behind the sample are shown in Fig.~\ref{fig:fields}(a) for an incident peak pump field of $E_0=115\kVcm$, with the corresponding intensity spectrum and spectrogram in (b) and (c), respectively (the spectrogram calculated with a temporal gate $w(t)=e^{-2t^2/T_g^2}$ with $T_g=10$~ps).
Here, the odd overtones from $n=3$ to 9 are clearly resolved.  
As shown in Fig.~\ref{fig:fields}(b), the relative spectral intensity vs. harmonic $n$ is quite close to that expected for an ideal nonlinear HG process ($I_n\sim a^{-n}$, dashed curve), as is the progressive reduction in pulse duration $T_n$ for each harmonic in Fig.~\ref{fig:fields}(c), where some deviations are expected due to saturation/propagation effects discussed below.

The theoretical modelling of the HG emission is based on Monte-Carlo (MC) simulations of an ensemble of holes in the time-dependent field. 
The simulations include the dynamics of the  
heavy-hole (hh), light-hole (lh) and split-off (so) valence bands, realistic 3D band structures, and acoustic-/optical-phonon scattering (as well as charged impurity scattering, when expected to contribute significantly), similar to the treatment employed previously in the context of high-frequency Si devices \cite{Jung1999,Jung2003}. 
As significant propagation effects also occur, including standing-wave effects for the pump field \cite{Meng22}, one must go beyond a description of the local response to describe the experimental results. 
Hence we embedded the MC simulations in a 1D (plane-wave) finite-difference time-domain (FDTD) propagation scheme 
with a full, self-consistent approach for the charges/fields for the simulations at high fields (as opposed to an earlier perturbative approach \cite{Jung2022}), typically with a step size of $\delta z=1\mum$ along the propagation direction. To investigate the microscopic HG mechanisms, local MC simulations were performed, i.e., at a single internal $z$-point in the bulk, as in our previous report \cite{Meng22}. 

An example of the emitted spectrum/spectrogram from such MC-FDTD simulations is shown in Fig.~\ref{fig:fields}(b) and (d), respectively, and shows very good agreement with the experimental data.  Note that the noise floor of the MC-FDTD results is due to the statistical fluctuations for the ensemble, which was restricted to $\leq 10^7$ holes in order to achieve practical calculation times.
As discussed below, these results contain both local-saturation and propagation effects, involving changes in the carrier distribution and scattering dynamics during the pump pulse.

\begin{figure}[!t]
	\includegraphics[keepaspectratio,width=0.5\textwidth]{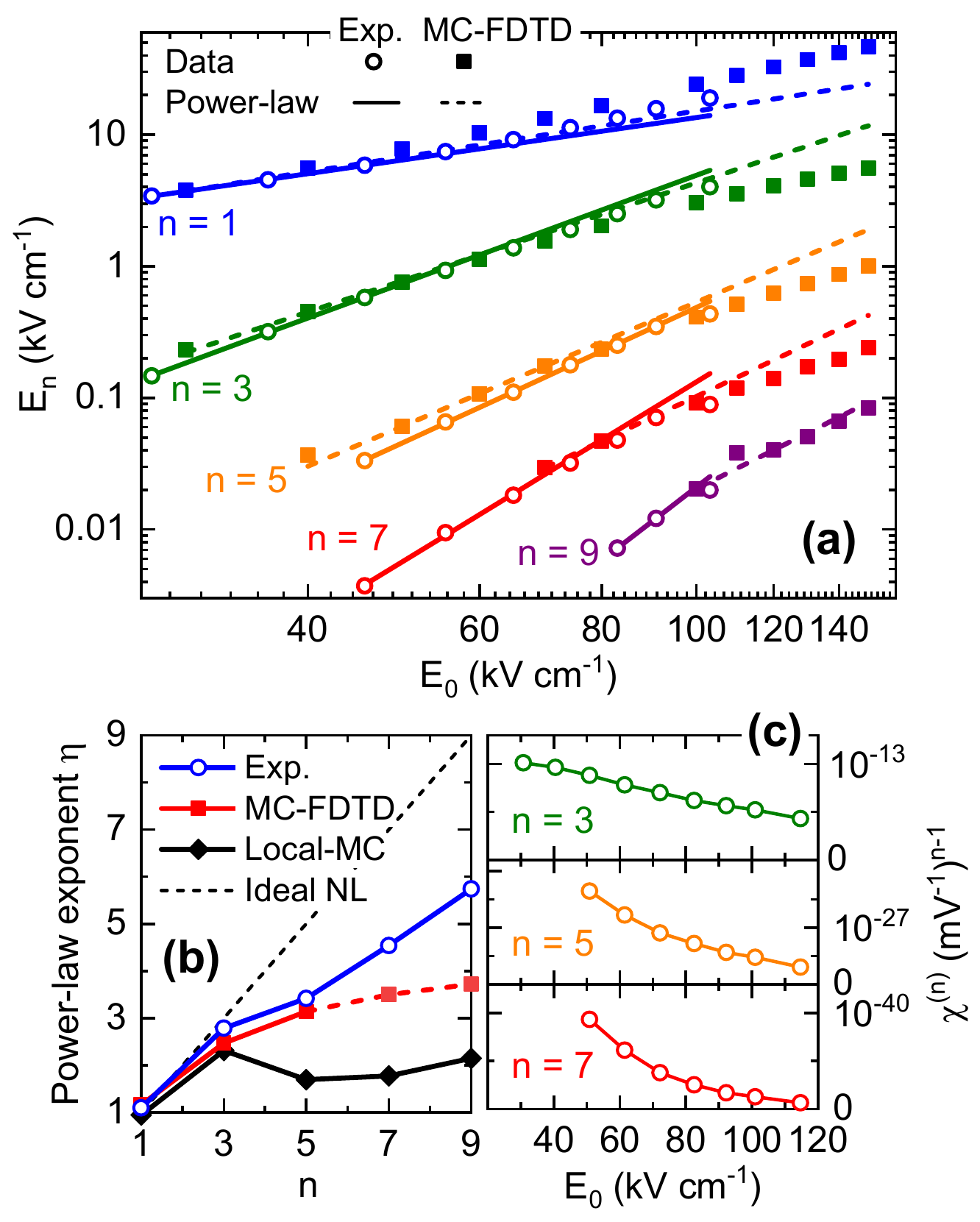}
	\caption{\label{fig:pumpdep}
	(a) Pump-field dependence of transmitted harmonic peak fields ($N_d=5.75\cdot10^{15}\pcmc$, $T=300$~K, $L=900$~µm): Experimental results (open circles), and MC-FDTD results (filled squares) for each odd harmonic $n=1-9$. Respective power-law fits included as straight lines.  To aid comparison, experimental peak pump fields ($E_0$, incident  external to the sample) scaled by a factor 0.9, while higher-harmonic fields are scaled by 1.7 ($n=5$), 2.0 ($n=7$), and 2.5 ($n=9$).
	(b) Power-law exponents from (a) vs. $n$, from experimental and MC-FDTD data, as well as single-point (bulk) MC simulations.  Note that the resolvable range of experimental data for $n=7,9$ are already well in the saturation regime.
	(c) Nonlinear susceptibilities $\chin{n}$ calculated from experimental peak fields in (a) after correction for Drude absorption (see main text).}
	\end{figure}

The dependence of the amplitudes of the transmitted harmonic fields on that of the incident pump field is shown in Fig.~\ref{fig:pumpdep}(a), for both experimental and MC-FDTD results. The data are obtained via band-pass filtering the spectral fields about each $\omega_n=2\pi\nu_n$ and extracting the peak temporal field after transforming each back to the time-domain (see Supplementary for temporal waveforms vs. $n$).
In order to aid visual comparison, the experimental pump field $E_0$ is scaled down by a factor of 0.9, while for $n=5,7,9$ modest correction factors are applied to the field values $E_n$ (as listed in the caption). A comparison of the unscaled experimental and simulated results is included in the Supplementary.  Given the experimental error margin for the field calibration, the agreement is seen to be very reasonable, especially as no scaling is applied to $E_{1,3}$.
Also included are power-law fits $E_n\sim E_0^{\eta_n}$ using the lowest available field ranges in each case; the exponent $\eta_n$ is plotted in Fig.~\ref{fig:pumpdep}(b).
Several key aspects are evident: Firstly, one sees with increasing field that the transmitted fundamental wave ($n=1$) grows super-linearly, in both the experimental and MC-FDTD data.
As discussed further below, this is due to the increase of the scattering rate of the holes at higher fields, which causes a reduction of the Drude absorption at $\nu_1$, and amounts to a bleach factor of ${\sim}2$ at the highest field (the linear Drude absorption depth of the sample at $\nu_1$ is 230\mum).
For $n=3$, one observes an initial growth close to the ideal power-law exponent of $\eta_3=3$, although the data visibly \textit{saturate} by a factor ${\sim}2$ at the highest fields, despite the bleach of the fundamental pump field.
A similar saturation is observed for $n=5,7,9$ (note that the noise levels in the MC-FDTD preclude fitting the data for $n=9$ at sufficiently low fields to avoid saturation).
These non-ideal field dependencies are due to an interplay between local (microscopic) saturation and propagation effects, and will be disentangled below with the aid of local MC simulations.
On the basis of the experimental fields, one can estimate an effective nonlinear susceptibility \chin{n} for each harmonic, the results of which are shown in Fig.~\ref{fig:pumpdep}(c) for $n=3,5,7$. 
These are calculated following standard approaches \cite{Hafe18} for uni-directional propagation (neglecting absorption and phase-matching)
which yields $E_n=\beta_n\chin{n}E_1^n$ with $\beta_n=i\omega_n L/(2^n n_r c)A_n$ ($n_r$ the refractive index, $c$ the vacuum speed of light). 
Due to the significant Drude absorption for both the pump and harmonics (which would lead to a underestimate of \chin{n} from the data), we include the factors $A_n=(1-e^{-a_nL}/a_nL)e^{-\alpha_nL/2}$ ($a_n=(n\alpha_1-\alpha_n)/2$) where $\alpha_n$ is the intensity absorption coefficient for each harmonic (the experimental bleach factor is also included to correct $\alpha_1$ to avoid overestimating \chin{n}, but only leads to a small correction, see Supplementary).
Clearly, the saturation effects with increasing $E_0$ lead to a reduction in the extracted values. For the lowest fields, we have $\chin{3}=1.0\cdot 10^{-13} \mVx{2}$, which is highly consistent with the value $\chin{3}=0.9\cdot 10^{-13} \mVx{2}$ one can deduce from previous reports for Si:B with nanosecond pulses at $\nu_1=610$~GHz \cite{Maye86} for this dopant density. 

We note that besides the pump-field dependence of the emission field strength, we also extracted the relative phases of the fields.  As described in the Supplementary, these vary by a small fraction of the fundamental cycle ($<0.2\pi$) while the trends vs. $E_0$ are also reasonably well reproduced in the MC-FDTD results. A comparison with the local-MC response shows these trends comprise both contributions from the local intra-cycle phase as well as a dominant contribution from the accumulated phase shifts due to propagation.

\begin{figure}[!t]
\includegraphics[keepaspectratio, width=0.5\textwidth]{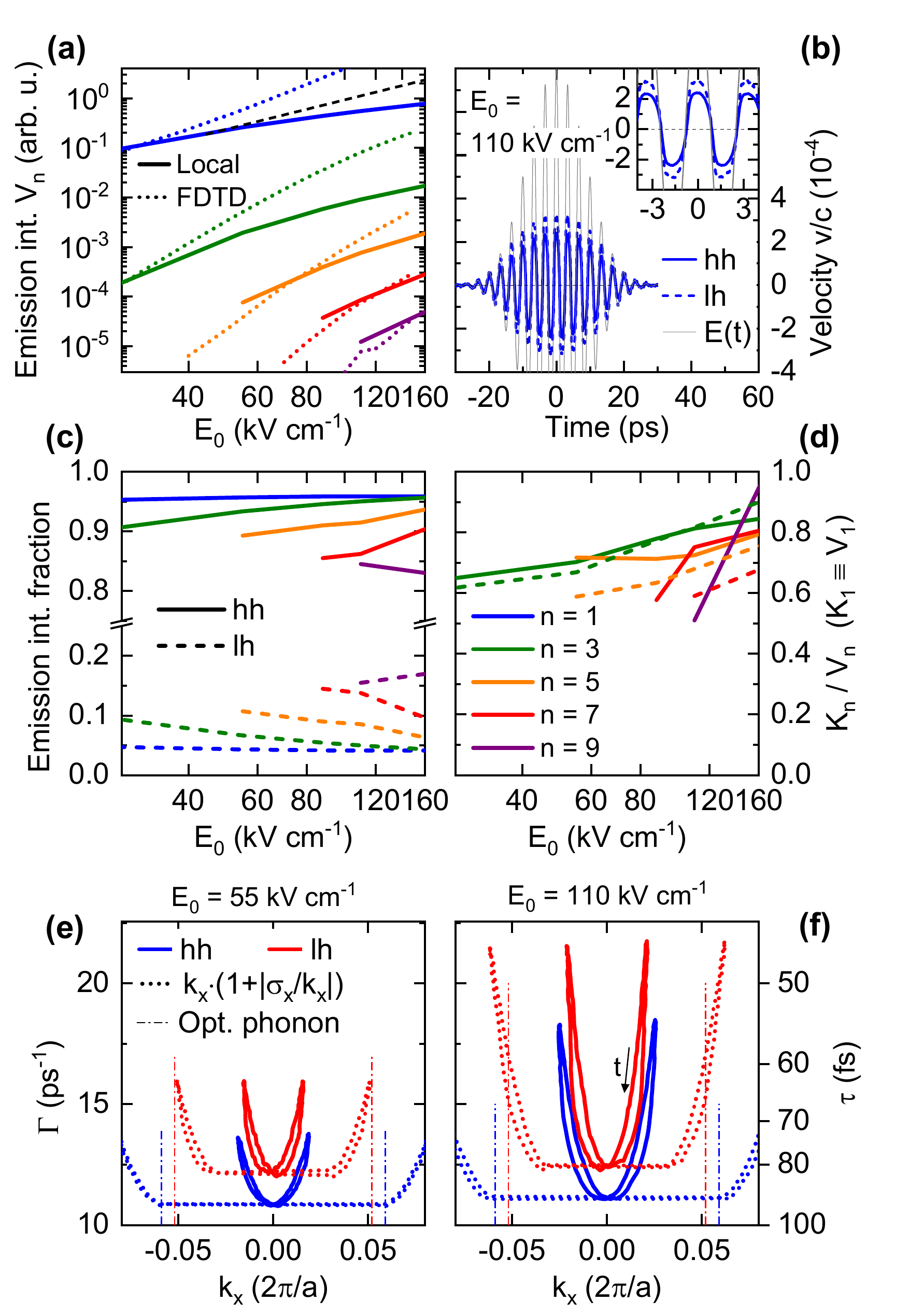}\caption{\label{fig:bulkmc} 
Local MC results ($N_d=5.75\cdot10^{15}\pcmc$, $T=300$~K). (a) Pump-field dependence of relative emission intensities $V_n$ for each odd harmonic $n=1-9$ (solid curves, legend in (d)). Scaled fluence from MC-FDTD (dotted curves) included for comparison (corresponding to data in Fig.~\ref{fig:pumpdep}(a)). Black dashed curve corresponds to a linear dependence, for comparison with the curves for $n=1$.
 Note that local internal pump fields used are $E'_0=\hat{t}_1 E_0$ (where $\hat{t}_1=0.46$ is the Fresnel field transmission entering the sample). %
(b) Time-domain (ensemble-average) velocity for $E_0=110\kVcm$ for both hh and lh, as well as scaled pump electric field profile $E(t)$. %
(c) Respective fraction of emission intensities from hh/lh for each $n$. %
(d) Ratio of nonlinear intensity due to $k_b(t)$ only ($K_{bn}$, see text) relative to emission intensity $V_{bn}$. %
(e, f) Scattering rates $S(t)$ during the peak cycle of $E(t)$ for two peak pump fields ($E_0=55, 110\kVcm$, respectively) plotted as loci vs. the ensemble averages $k_{bx}(t)$.  Also included are the same loci, but vs. $k_{bx}\cdot(1+|\sigma_{bx}/k_{bx}|)$ to reflect the extent of the high energy tail of the distributions.  Thresholds for opt. phonon emission for each band (corresponding to an energy of $\Eopph=63.3$~meV) included as vertical lines.  Orientation of temporal hysteresis indicated by arrow in (f).
}
\end{figure}

In Fig.~\ref{fig:bulkmc}, we plot a selection of results from local (single $z$-point) MC simulations vs. field, where we stored and ana\-ly\-zed the time-dependent, ensemble-averaged band velocities $v_b(t)$ ($b=\hho,\lho,\soho$) and  occupation-weighted total velocity $v(t)$, wavevectors  $k_{bj}(t)$ and their rms spreads $\sigma_{bj}(t)$ ($j=x,y,z$, where the pump field is polarized along the $x$-direction), band populations $N_b(t)$ and scattering rates $\Gamma_b(t)$. 
As the so-band has only a very minor contribution, it was omitted in the following analyses.
In Fig.~\ref{fig:bulkmc}(a), we plot the relative emission intensities $V_n$ for each harmonic $n$, obtained by integrating the spectral intensity of $V(\omega)=|\mathcal{F}\{v(t)\}|^2$ about each $\omega_n$.  Also included are the fluences $F_n\propto|E_n|^2$ from the MC-FDTD results in Fig.~\ref{fig:pumpdep}(a) (scaled for comparison by a common constant for all $n$), which in the absence of any propagation effects would coincide with $V_n$ for $n>1$.
One observes saturation in the local response with increasing field for all overtones as well as the fundamental. For the latter, its origin is clearly demonstrated in Fig.~\ref{fig:bulkmc}(b), where we plot $v(t)$ for hh and lh with a scaled profile of the pump field $E(t)$. The inset displays three oscillation periods at the center of the pulse. One sees how $v(t)$ undergoes strong clipping at each peak of $E(t)$, which is due to a rapid rise in momentum scattering, discussed further below.  
Integrated along the beam path, the decrease in local emission intensity is, however, overcompensated by reduced loss as shown by the blue lines (full and dotted) in Fig.~\ref{fig:bulkmc}(a).
The loss reduction is also a consequence of the increase in scattering rate for the local current which manifests as a weaker Drude absorption at $\omega_1$ during propagation, and leads to the bleach effect seen in Fig.~\ref{fig:pumpdep}(a).
Comparing the local and FDTD results in Fig.~\ref{fig:bulkmc}(a) allows one to draw additional conclusions concerning the role of propagation for the harmonics $n>1$: The local saturation in the emission intensity for $n=3$ is compensated by the bleach of the Drude absorption during propagation, weakening the saturation effect for the transmitted signal.  This effect is also present to a lesser degree for $n=5$, but is essentially absent for $n=7,9$, as here the increased scattering rate no longer causes a significant bleach of the Drude absorption.  Moreover, with increasing $n$, the effects of a finite phase-mismatch $\Delta k_n\sim \omega_n(n_{rn}-n_{r1})/c$ should also lead to additional saturation effects during propagation.
In Fig.~\ref{fig:bulkmc}(c), we compare the relative contributions to the total emission intensity from hh and lh for each $n$. As expected, the linear response ($n=1$) is dominated by hh, due to their much higher occupation, $N_{\hho}/N_0\approx 0.85$ in thermal equilibrium, which does not vary significantly during the pump excitation (in contrast to our previous simulations at $T=10$~K \cite{Meng22} where $N_{\hho}$ rose to $0.93$ in non-equilibrium, further depleting $N_{\lho}$).  For the overtones, the relative contribution from lh increases with $n$, indicating that they have a stronger nonlinear response per-hole than the hh (see below).

As mentioned in Sec.~\ref{sec:overview}, one can distinguish two main mechanisms for the HG process: band non-parabolicity (``\banh") and energy-dependent scattering (``\edsr").
To gauge their relative roles, we employ the same approach we used previously (Ref.~[\onlinecite{Meng22}], Fig.~4), i.e., to compare the emission intensity $V_{bn}$ for each band with the spectral intensity calculated from the occupation-weighted wavevector component $k_{bx}(t)$, i.e., $K_{bn}$ obtained from integrating $K_b(\omega)=|\mathcal{F}\{k_{bx}(t)\}|^2$ about each $\omega_n$ (in contrast, we determined $K_{b}(\omega)$ in Ref.~[\onlinecite{Meng22}] from a 1D$\langle k \rangle$ calculation for a rigid 1D wave packet instead of Monte Carlo simulations for an ensemble of holes). 
While the band non-parabolicity 
indeed has a finite influence on the precise scattering processes and hence $k_{bx}(t)$, the quantity $K_{bn}$ still provides a reasonable measure for the nonlinear response in the hypothetical absence of \banh.  By normalising $K_{b1}\rightarrow V_{b1}$, for the overtones a ratio of $K_{bn}/V_{bn}\rightarrow$0 or 1 corresponds to pure \banh{} or \edsr, respectively.
As shown in Fig.~\ref{fig:bulkmc}(d), the $K_{bn}/V_{bn}$ values are all above 0.5 in the field ranges where harmonics can be extracted, increasing toward unity as the field amplitude increases. It indicates that, for the highest electric field used in our experiment, \edsr{} dominates. One could also envisage that, at the electric field below $ 50\kVcm$, the \banh{} also plays an important role for certain higher order harmonics, i.e $n=7,9$. This is in stark contrast to our earlier study~\cite{Meng22} for $n=3,5$ at $T=10$~K with $N_d\leq 10^{14}\pcmc$ and somewhat lower fields ($E'_0\lesssim 25\kVcm$, $E_0\lesssim 50\kVcm$), where $K_{bn}/V_{bn}\ll 0.01$, and hence \banh{} heavily dominated the HG response.  The major reason for this is, that the higher fields in the present study allow to reach the threshold for optical phonon emission (the TO/LO phonon being degenerate at the $\Gamma$ point with $\Eopph=63.3$~meV \cite{Dargys1996}). This is not immediately obvious in Fig.~\ref{fig:bulkmc}(d) where $K_{bn}/V_{bn}$ for $n=3$ does not fall rapidly as the field strength decreases to  $E_0\lesssim 50\kVcm$, 
the field range of our previous study \cite{Meng22}.  This can be traced to the broadening effect by the higher value of $T$ (300~K here, 10~K in [\onlinecite{Meng22}]), whereby the broadened  Fermi-Dirac distribution brings the more energetic carriers already close to \Eopph{} in thermal equilibrium before the pump excitation (the band-filling at these densities $N_d$ is still only a minor effect).
This suppresses any sharp threshold behavior for $K_{bn}/V_{bn}$ vs. $E_0$ here, although experiments at somewhat lower $T$ (but sufficiently high to maintain a thermally ionized hole population) should show a much clearer transition. 

In Fig.~\ref{fig:bulkmc}(e,f) we plot the loci of the instantaneous, ensemble-averaged scattering rates $S(t)$ vs $k_{bx}(t)$ (for a cycle at the peak of the pump pulse) for peak pump fields $E_0=55$ and $110\kVcm$, respectively.
To better reflect the magnitude of $k_{bx}$ reached by the high-energy tail of the carrier distribution, we also plot the data vs. $k_{bx}(t)\cdot(1+|\sigma_x(t)/k_{bx}(t)|)$, i.e., $k_{bx}(t)$ extended by the instantaneous rms width $\sigma_x(t)$ of the distribution.
Here one clearly sees that the $k$-space extent of the carrier distributions is significantly broader than the ensemble average.
At $k_x=0$, the hh distribution (unlike the lh distribution) already touches the threshold for LO/TO phonon emission, although the hh and lh scattering rates are still dominated by acoustic phonon scattering. When driven out in $k$-space, the hh and lh populations experience a strong increase of the scattering rate, much of which appears to be due to reaching and crossing the threshold for LO/TO phonon emission.
At $E_0=110\kVcm$, the increase in scattering is steeper for lh than hh. Moreover, a comparison of $K_{bn}$ vs. $E_0$ (not shown) also shows a higher ratio $K_{\lho,n}$/$K_{\lho,1}$ compared to $K_{\hho,n}$/$K_{\hho,1}$, i.e., the lh nonlinear response is dominantly from the higher \edsr, and not a higher \banh.
These effects arise due to the fact that the lh initially absorb more kinetic energy from the pump field, allowing the highest-energy lh to experience stronger optical phonon emission before the lh$\rightarrow$hh intra-cycle energy equilibration takes full effect.

\subsection{Low-temperature harmonic generation}\label{sec:hg4K}

We turn now to the low-temperature case, where the vast majority of acceptors are in the neutral state and only a small residual density of holes reside in the valence bands before the pump pulse excitation.
In our previous study of Si:B at low temperature \cite{Meng22}, due to both the lower peak pump fields ($E_0\lesssim 50\kVcm$) and higher frequency ($\nu_1=1.29$~THz), we did not reach the threshold for significant tunnel ionization. Nevertheless, we could resolve both 3HG and 5HG in the emitted fields (measured only as intensity spectra obtained with far-infrared Fourier transform spectrometry), which arose from the residual band-hole density (which was seen to grow with increasing field due to carrier multiplication during the pump pulse, resulting in a moderately field-dependent value of \chin{3,5}).  Moreover, as mentioned above, the nonlinearity giving rise to the HG was found to be dominated by \banh, in particular as most holes did not reach the threshold energy for optical phonon emission whereby \edsr{} should become very strong.

Here at low temperature, we can reach peak external incident pump fields of $E_0{\sim}80\kVcm$ (and internal fields $E'_0{\sim}40\kVcm$), and more decisively at a significantly lower frequency of $\nu_1=0.3$~THz, such that the ponderomotive energy $U_{p,b}=q^2E_0^2/(4m_b \omega_1^2)$ \cite{Lev94} will be up to a factor ${\sim}50$ higher ($U_{p,\lho}\sim 1.3$~eV for lh with $m_\lho/m_e=0.15$), and the Keldysh parameter (an inverse measure of the onset of tunnel ionization for a required ionization energy $I_p$) $\gamma_b=\sqrt{I_p/2U_{p,b}}$ also 7 times higher ($\gamma_\lho\sim 0.13$, well into the tunnel regime).  Note we tentatively employ these expressions from the gas phase, where scattering (dephasing) is neglected and one assumes a ballistic trajectory for the ionized charges, as addressed 
below.

\begin{figure}[!t]
	\includegraphics[keepaspectratio,width=0.5\textwidth]{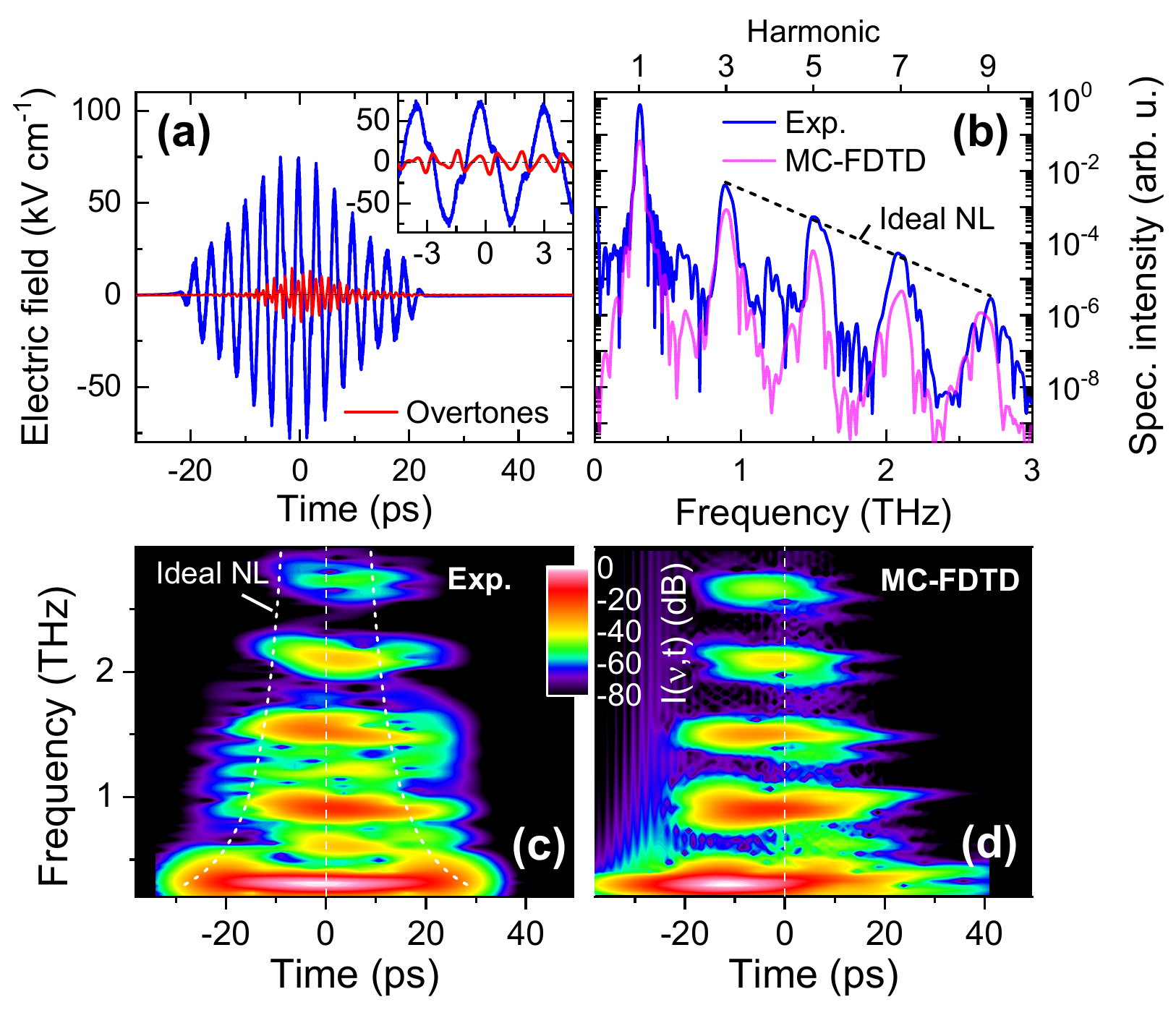}
	\caption{\label{fig:fields4K} (a) Experimental transmitted temporal field for Si:B sample ($N_d=5.0\cdot10^{16}\pcmc$, $T=4$~K) for an incident pump peak field of $E_0=81\kVcm$ (blue curve), as well 
	as the field of the overtones without fundamental (red).  Inset shows detail around the pulse peak.  Note that a temporal window was applied to the wings of the pulse to suppress noise and reflections in the sample.
	(b) Corresponding intensity spectrum, as well as that from MC-FDTD simulations including tunnel ionization for $E_0=80\kVcm$ ($N_d=5.9\cdot10^{16}\pcmc$). Both spectra plotted with the same absolute intensity scale.
	(c) Experimental and (d) MC-FDTD spectrograms. Additional annotations per Fig.~\ref{fig:fields}.
	}
\end{figure}

As per the results in Fig.~\ref{fig:fields} for $T=300$~K, we first show a summary of the experimental emitted fields in Fig.~\ref{fig:fields4K} for the highest pump field $E_0=81\kVcm$.
Compared to the results for $T=300$~K, the bandwidths of the overtones here are larger (see also Fig.~\ref{fig:overview}(c)), increasingly so for higher $n$, and the intensity ratios between successive overtones  are significantly smaller.  The latter is also evident in the time-domain overtone field (Fig.~\ref{fig:fields4K}(a), red curve) which shows a more complex interference of the $n=3$ wave with those for $n>3$.
The larger bandwidths could in principle indicate either successively shorter harmonic pulses (decreasing faster than the $T_n\sim T_1/\sqrt{n}$ rate for an ideal nonlinearity with a Gaussian pulse profile) or a frequency modulation of each HG field: An inspection of the experimental spectrogram (Fig.~\ref{fig:fields4K}(c)) shows that the latter effect is significant, which we attribute to the fact that the hole populations vary during the pulse due to the tunnel ionization (addressed further below), such that the wave-mixing is more complex than just a multi-photon convolution of the components of the pump pulse spectrum. %
That the broadening is mostly due to local effects (rather than nonlinear refraction during propagation) is confirmed by comparing the spectra from FDTD-MC and local-MC simulations (presented below), the latter also showing such increased spectral widths.

As in Sec.~\ref{sec:hg300K}, we performed MC simulations of the HG experiments. This required extending the MC scheme to incorporate the tunnel ionization process injecting holes into the bands, with a time-dependent density $N_b(t)$.
To calculate the ionization rate $\partial_t N_b(t)$, we employed the ionization probability rates $w_{i,b}(E)$ established in the literature for a static electric field \cite{Darg95,Nie84} and integrate $\partial_t N_b = w_{i,b}(t)(N_d-N)$ 
($N=\Sigma_b N_b$) during each time step with the field $E(t)$.
The use of a static-field model for the instantaneous tunnel ionization rate has been shown to hold reasonably well in the context of ionization of gas atoms/molecules, at least in the strong-field regime ($\gamma\ll 1$) \cite{Yudin01,Reiss08,Boro22}.
As asserted in \cite{Nie84}, the tunneling vs. field should be dominated by ionization into the lh band, which seems reasonable due to the $U_{p,b}\propto 1/m_b$ dependence for the ponderomotive energy.  
As a simplified approach, we then take $w_{i,\hho}\rightarrow 0$ and inject holes only into the lh band, which then can rapidly scatter into the hh band, as modeled by the MC treatment.
Details of the ionization rate (formula, parameters for Si:B and plot of $w_{i,\lho}(E))$ are given in the Supplementary.

\begin{figure}[!t]
	\includegraphics[keepaspectratio,width=0.5\textwidth]{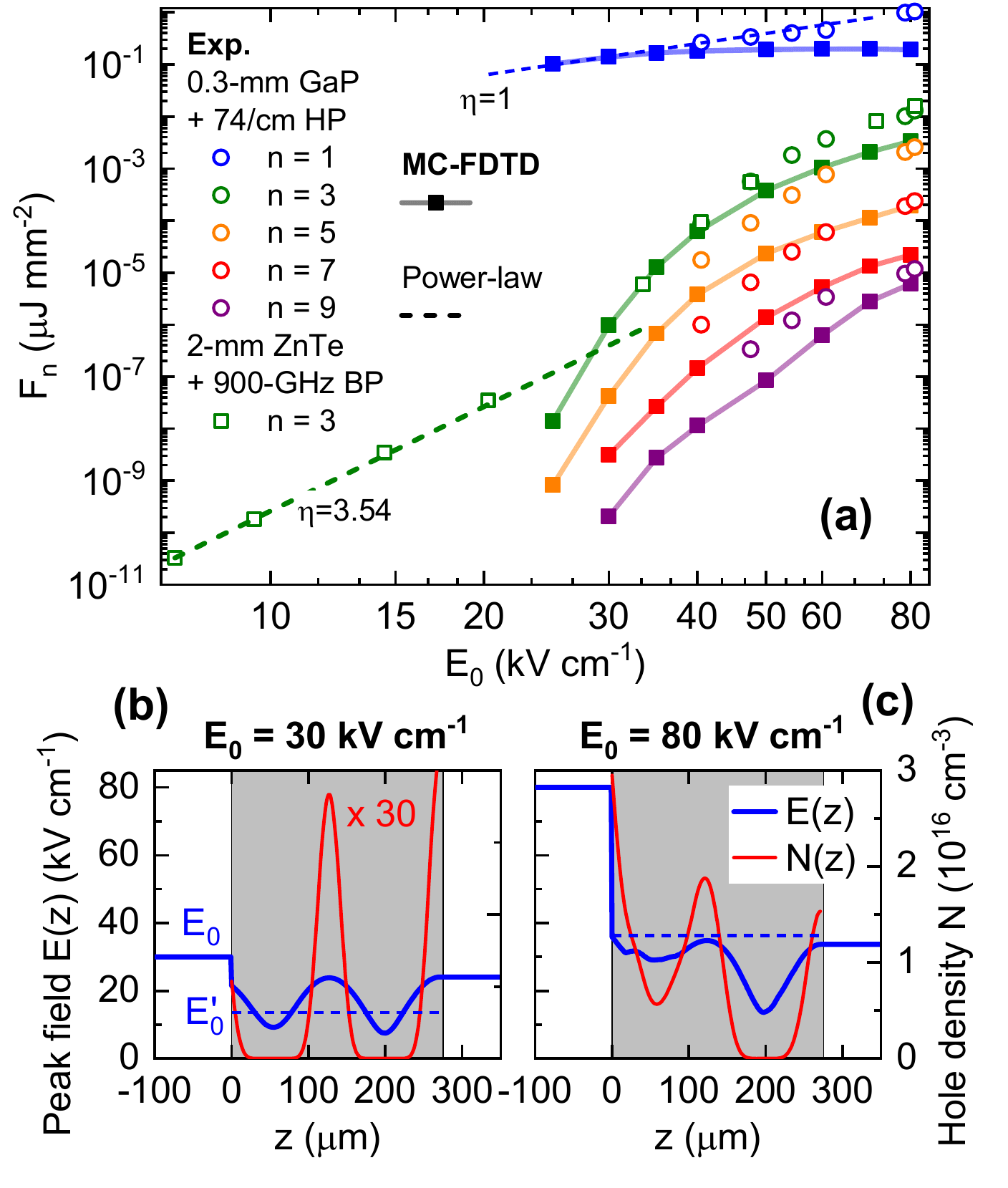}
	\caption{\label{fig:pumpdep4K} (a) Pump-field dependence of transmitted harmonic fluence ($N_d=5.0\cdot10^{16}\pcmc$, $T=4$~K, $L=272$~µm): Experimental (open circles, with additional higher sensitivity measurements for $n=3$ as open squares), and MC-FDTD results (filled squares, using $N_d=5.9\cdot10^{16}\pcmc$) for each odd harmonic $n=1-9$.  Power-law fits for $n=1$ and $n=3$ included as straight dashed lines (exponents $\eta$ as indicated).
	(b,c) Spatial profile of peak electric field $E(t)$ and ionized hole density $N(z)$ for two values of pump field $E_0=30$ and $80\kVcm$, respectively.  Only incident field amplitude $E_0$ (without reflected field) shown for $z<0$.  Also included are the nominal internal pump fields $E'_0=\hat{t_1}E_0$ (used for the local MC simulations when specifying $E_0$).
	}
	\end{figure}

The corresponding intensity spectrum and spectrogram from MC-FDTD calculations for conditions close to the experimental ones are shown in Fig.~\ref{fig:fields4K}(b) and (d), respectively.  The simulated results do exhibit a reasonable qualitative correspondence to the experimental data, although certain systematic deviations are present, as discussed in the following.  
We first compare the dependence of the HG emission vs. pump field in Fig.~\ref{fig:pumpdep4K}, which includes both experimental and MC-FDTD data.
As the broadband EO field detection 
was only sensitive enough to resolve the harmonics for $E_0 \gtrsim 40\kVcm$, we augmented the measurements of the 3HG-emission with a more sensitive low-bandwidth EO sensor for pump fields down to $E_0{\sim}10\kVcm$, as shown.
As significant distortions are seen in the envelopes of the filtered time-domain harmonics (see Supplementary), we choose here to plot the results in terms of the emitted fluence ($F_n=\epsilon_0 c\int{dt E^2_n(t)}$) to avoid any artifacts. %
One sees that at low pump fields, the 3HG emission closely follows a power-law dependence, $F_n\propto E_0^{2\eta_3}$ with $\eta_3=3.54$.
This behavior is comparable to that seen in our previous study at $T=10$~K (with the same B dopant concentration as here) \cite{Meng22}, where a value of $\eta_3=4.2$ was determined, which exceeds the value $\eta_3=3$ (for an ideal nonlinear process in a static medium)
due to field-driven multiplication \cite{Gani86} of the residual band carriers (density $N_r$) during the pump pulse.
The different value of $\eta_3=3.54$ here can be attributed to the different pump frequency.
In \cite{Meng22, Meng2020}, we employed the impact ionization model from \cite{Gani86} in its high-field limit, where the carrier multiplication factor $f=\Delta N/N_r\sim 1-E_m^2/E^2$, with the characteristic field constant $E_m$ increasing with $\omega_1$. 
Hence $\partial_E f \sim +2E_m^2/E^3$, which should be larger for the previous experiments with $\nu_1=1.29$~THz and hence produce a larger value of $\eta_3$, which is at least qualitatively consistent with the two results.

Turning now to the higher field range $E_0\gtrsim 25\kVcm$, one sees that the experimental 3HG emission grows rapidly (by more than 4 orders of magnitude in fluence) with a reasonably sharp onset, accompanied by the higher harmonics $n=5,7,9$ almost in proportion.  Indeed, the MC-FDTD simulations also show a similar pump-field dependence, which is due to the onset of tunnel ionization and resultant nonlinear response of the ionized band holes, which then begins to saturate due to similar effects seen above for $T=300$~K. Considering that no rescaling is applied to the results, the agreement is quite remarkable.

The importance of employing a full FDTD treatment for modeling the experimental data is demonstrated in Fig.~\ref{fig:pumpdep4K}(b,c), where we plot the spatial distributions of both the peak field $E(z)$ (i.e., the maximum amplitude during FDTD propagation) and photoionized hole density (after propagation, $N(z)=\Sigma_b N_b(z)$) for a value of $E_0$ at the onset of tunnel ionization (Fig.~\ref{fig:pumpdep4K}(b)) and above (Fig.~\ref{fig:pumpdep4K}(c)). In Fig.~\ref{fig:pumpdep4K}(b), one clearly sees the standing-wave profile in $E(z)$ due to multiple reflections of the field in the sample, which manifests as three field-enhancement peaks (at each surface and in the center of the Si:B sample), which exceed the field $E'_0$ one would have if only accounting for the Fresnel transmission coefficient $\hat{t}_1$ of the incident field $E_0$ -- this will be addressed again below in assessing the local-MC results.  (Note that these effects are not significant for the $T=300$~K case above, due to the strong, pre-existing Drude absorption from the thermally ionized carriers.)
This standing-wave effect also gives rise to a strong dependence of the photoionized hole profile $N(z)$, and for the case in (b) where one is still close to the initial, exponential onset of ionization, the small asymmetry in $E(z)$ manifests as a significant asymmetry in $N(z)$.  For the higher-field case in Fig.~\ref{fig:pumpdep4K}(c), one sees that the field profile becomes more complex, affected by depletion and propagation effects, leading to a reshaping of the standing-wave pattern in $N(z)$.  
Moreover, the field enhancement is actually suppressed and one sees peak fields closer to the nominal value $E'_0$.
Clearly a neglect of these effects (e.g. using a simplified, single-pass uni-directional propagation model, or a bulk layer -- see below) would have a significant impact on the predicted HG emission.

One discrepancy remains. It concerns the pump-induced Drude absorption of the fundamental predicted in the MC-FDTD results (see data for $n=1$ in Fig.~\ref{fig:pumpdep4K}(a)).
While the experimental transmitted fluence continues to grow linearly in the pump fluence (even showing a superlinear growth at the highest fields), for the MC-FDTD results one sees that this undergoes a significant saturation due to the Drude absorption by the ionized carriers.  This manifests as a suppression of the trailing half of the pump pulse exiting the sample, as seen in the spectrogram in Fig.~\ref{fig:fields4K}(d).
We performed several test simulations to look for any possible resolution to this discrepancy: (i) reduction of the dopant density, (ii) allowing for rapid recombination of ionized holes with their parent ions (both of which would reduce the Drude absorption at $\omega_1$) -- however, in all tests the agreement for the overtones was significantly degraded, with the predicted fluences $F_n$ falling further below the experimental levels. Moreover, we tested that injection of the holes rather into the hh band does not lead to any significant changes.
While the tunnel-ionization-rate model has not been tested in this regime in the literature, evidently the onset vs. pump field in the simulations is close to quantitative. One possible hypothesis is that fewer holes would be generated than the employed ionization model predicts, and that there is an additional contribution to the harmonics e.g. from recollision with their parent ions (not included in the $k$-space MC simulations). As presented in Sec.~\ref{sec:discussion}, this seems unlikely due to the strong intra-cycle scattering, which should strongly suppress any (coherent) recollision.

\begin{figure}[!t]
	\includegraphics[keepaspectratio,width=0.5\textwidth]{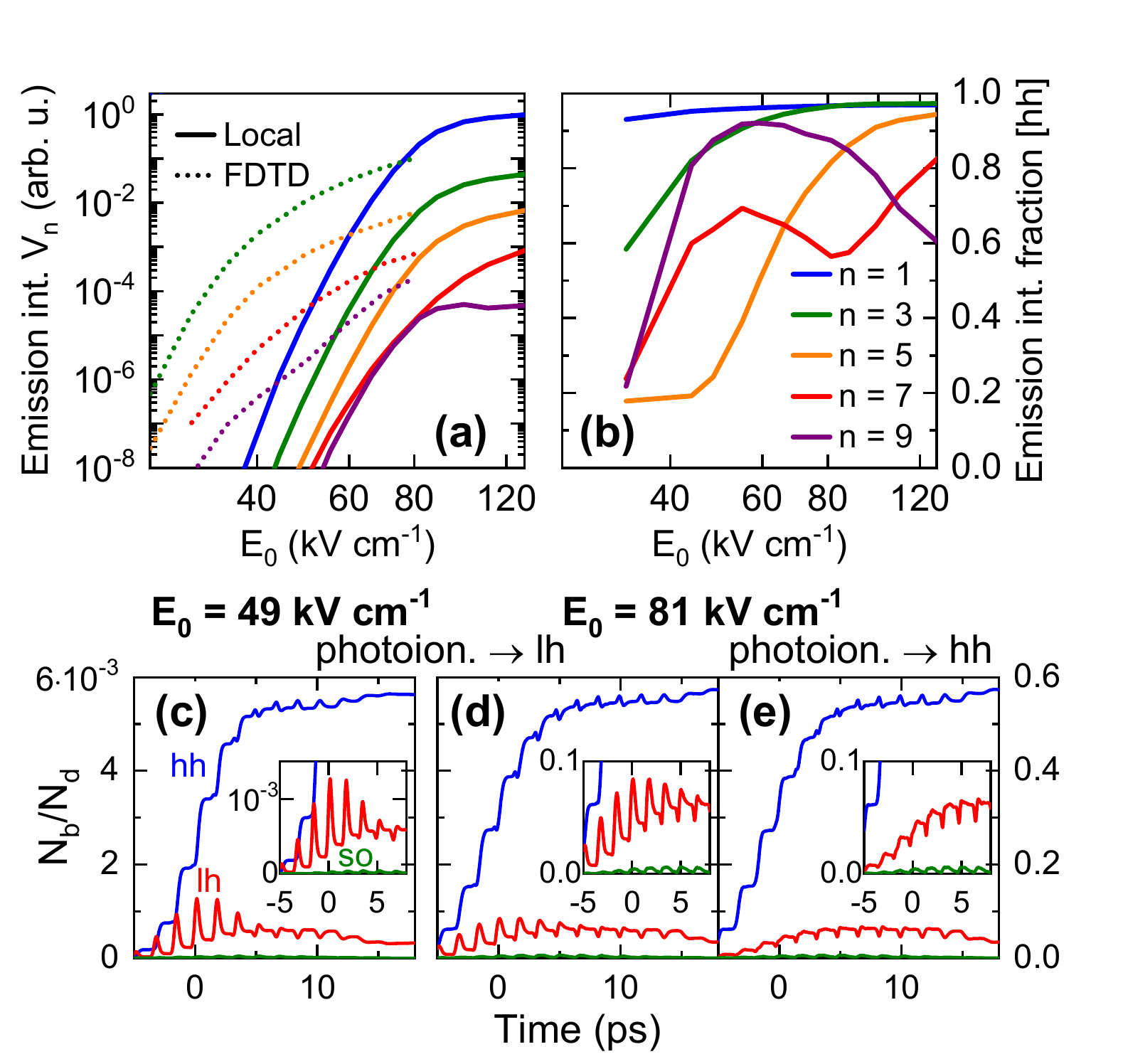}
	\caption{\label{fig:bulkmc4K} Local MC results ($N_d=5.9\cdot10^{16}\pcmc$, $T=4$~K). %
	(a) Pump-field dependence of relative emission intensities $V_n$ for each odd harmonic $n=1-9$ (solid curves, legend in (b)). Scaled fluence from MC-FDTD (data from Fig.~\ref{fig:pumpdep4K}(a), here dotted curves) included for comparison. As per Fig.~\ref{fig:bulkmc}, the field values $E_0$ given correspond to the external incident fields (see Fig.~\ref{fig:pumpdep4K}(b)). %
	(b) Fraction of emission intensities $r_{\hho,n} = V_{\hho,n}/(V_{\hho,n}+V_{\lho,n})$ from hh for each $n$ vs. $E_0$ ($r_{\lho,n}=1-r_{\hho,n}$ omitted for visual clarity). %
	(c,d) Time-dependent hole band populations (relative to $N_d$) for two pump fields $E_0=49\kVcm$ and $81\kVcm$, respectively.  Initial photoionized holes are taken to enter exclusively into the lh band -- see text. (e) As per (d), only using a simulation where initial photoionized holes enter exclusively into the hh band for comparison. Vertical scale for (c) at left, for (d,e) at right. Insets show magnified ranges.  
	}
\end{figure}

Proceeding on the basis that important aspects of the HG process are described by the MC simulations, we  inspect the local response, again using MC simulations at a single $z$-point, as summarized in Fig.~\ref{fig:bulkmc4K}.
In Fig.~\ref{fig:bulkmc4K}(a), we plot the local emission intensity vs. pump field for each harmonic.  Also included for comparison is the scaled fluence from the MC-FDTD simulations above.  One notices immediately that the threshold fields for the latter (governed in both cases by the onset of tunnel ionization) are significantly lower (by a factor ${\sim}2$).  This is due to the standing-wave field enhancement effect presented above (Fig.~\ref{fig:pumpdep4K}(b,c)).  With increasing pump fields, this field enhancement becomes increasingly suppressed, giving rise to a saturation behavior in the MC-FDTD (and, according to Fig.~\ref{fig:pumpdep4K}(a), also the experimental) fluence which is not due to the inherent local response.
One sees also that for the local response, the highest harmonics ($n=7,9$) emerge rather close to the onset of local saturation effects, which are due to similar effects for the $T=300$~K case in Sec.~\ref{sec:hg300K}, i.e., primarily the onset of the optical phonon emission.  However, as shown before in Fig.~\ref{fig:pumpdep4K}(a), the FDTD propagation effects (including phase mismatch which becomes more severe with increasing $n$) result in relative harmonic yields more consistent with the experimental data.

In Fig.~\ref{fig:bulkmc4K}(b) we show the relative emission intensity contribution from hh (the remaining fraction again being dominated by lh, with a negligible contribution from the so-band holes).  Compared to the $T=300$~K results above (Fig.~\ref{fig:bulkmc}(c), where the lh contribution was below 20\% for all harmonics) here for low pump fields, one sees that the lh contribution can actually \textit{dominate} the HG emission for certain $n$, with a complex dependence on $E_0$, although the hh contribution becomes strongest for all $n$ at the highest fields. 
To assess this intriguing result, in Fig.~\ref{fig:bulkmc4K}(c,d) we plot the local time-dependent hole populations $N_b(t)$ (relative to the dopant density taken as $N_d=5.9\cdot 10^{16}\pcmc$), for both low and higher fields, respectively.
As expected from the implementation of the photoionization model, at each intra-cycle peak of the pump field one sees a rapid growth in $N_\lho$.  Within the subsequent half-cycle, interband scattering drives the majority of the newly ionized holes into the hh band, such that during the pulse the relative densities approach those dictated by the density of states in the respective band.
This intra-cycle interband scattering effect is specific to the case with photoionization, while the modulation of the band populations with the thermally populated bands ($T=300$~K) is much weaker. 
As a control simulation, we also modified the MC treatment to rather inject all holes into the hh band, which results in the time-dependent band densities shown in Fig.~\ref{fig:bulkmc4K}(e).  In this case, one sees that the intra-cycle inter-band scattering is weaker, and (as expected) inverted between lh and hh.  Interestingly, at this high field of $E_0=81\kVcm$ ($E'_0=37\kVcm$), the predicted harmonic emission intensities are almost identical to the values for lh ionization, although the relative lh contribution decreases.
Given the highly non-linear intra-cycle kinetics of $N_b(t)$, in the following we inspect more closely how this might contribute to the HG emission.

\begin{figure}[!t]
	\includegraphics[keepaspectratio,width=0.5\textwidth]{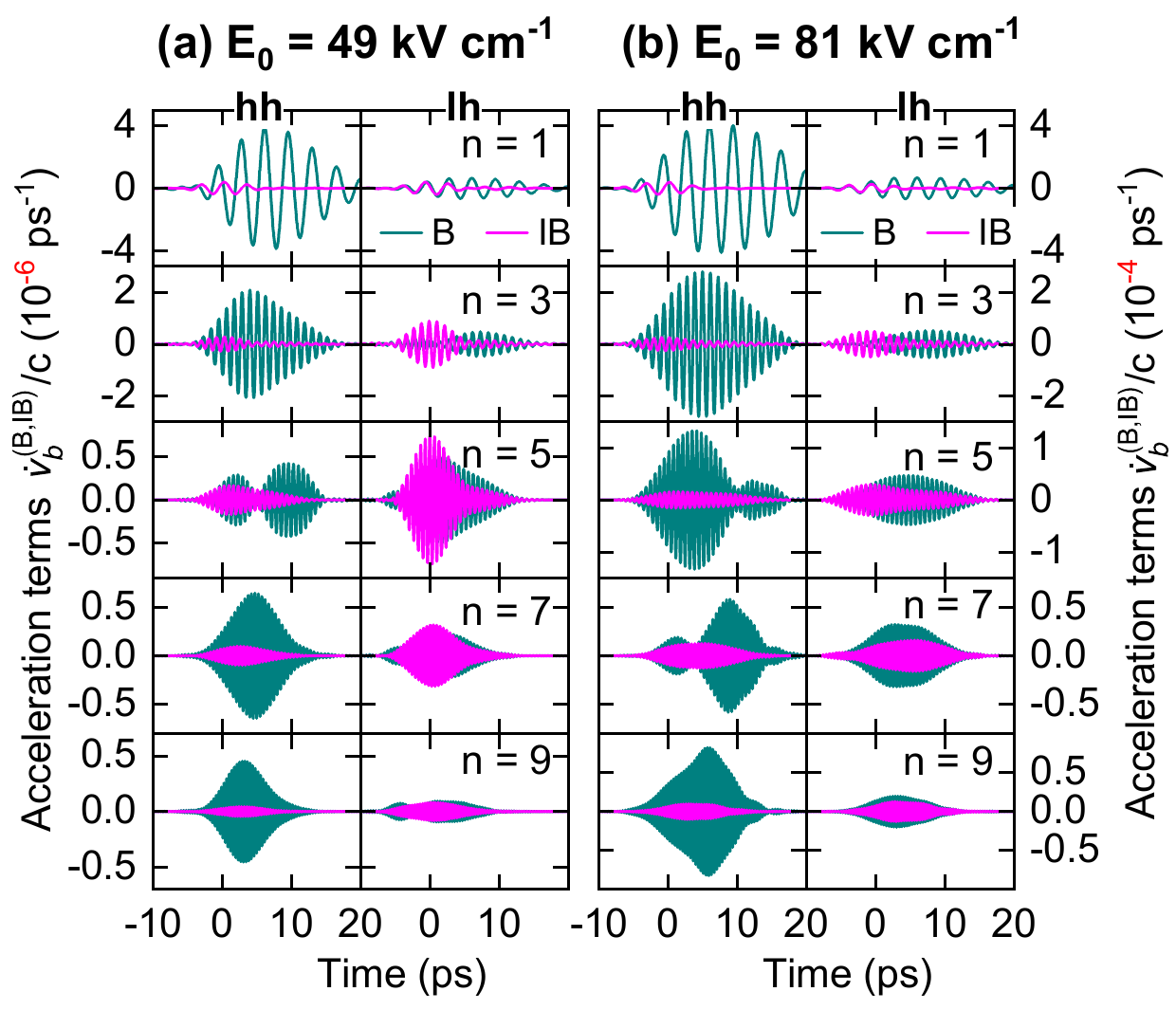}
	\caption{\label{fig:bulkmc4Kacc} 
	Local MC results ($N_d=5.9\cdot10^{16}\pcmc$, $T=4$~K): Decomposition of (occupation-weighted) hole acceleration into ``intraband'' (\accB{b}) and ``interband'' (\accIB{b}) contributions (see text for definitions) for (a) $E_0=49\kVcm$ and (b) $E_0=81\kVcm$, for each band ($b=\hho, \lho$), filtered for each harmonic $n$.  Note different vertical scaling in (a) and (b).
	}
\end{figure}

One approach to analyze the relative contributions to HG emission from motion in a given band (``intraband" contribution) with that due to band-population changes (``interband" contribution), is to calculate the (occupation-weighted) acceleration components, i.e., with 
$v(t)=\Sigma_b n_b(t)v_b(t)$ (where $n_b=N_b/N_d$ is the density of ionized holes in band $b$ relative to the dopant density) one obtains  
$\dot{v}=\partial_t v=\Sigma_b(\accB{b}+\accIB{b})$, where
$\accB{b}=n_b\dot{v}_b$ and $\accIB{b}=\dot{n}_b v_b$.
In this case, \accIB{b} comprises both the contributions from photoionization (``generational nonlinearity") and from $\hho \leftrightarrow \lho$ scattering. Although our physical situation differs from the well-studied HG process from dissociated e-h pairs \cite{vampa14,vampa15,banks13}, this is reasonably consistent with the terminology introduced there.
While this definition tacitly assumes that carriers entering a band adopt the ensemble average velocity in that band $v_b(t)$, in the absence of a more rigorous treatment one at least observes that $\accIB{b}$ vanishes for $\dot{N}_b(t)=0$, while the occupation weighting provides a direct measure of the relative contribution from each band.
By applying a band-pass filter about each harmonic frequency $\omega_n$, we can inspect their respective contributions to each harmonic.
This is presented in Fig.~\ref{fig:bulkmc4Kacc} for both a low (onset of photoionization) and high (saturation regime) pump fields for both hh and lh bands.
Beginning with the low-field case (Fig.~\ref{fig:bulkmc4Kacc}(a), $E_0=49\kVcm$), several features can be observed in the results.
Firstly, one sees that the lh contribution is significant, and even dominates the HG emission for $n=5$ (as seen in  Fig.~\ref{fig:bulkmc4K}(b) for this field value, noting that the emission intensity scales with $[v(t)]^2$).  Moreover, for the lh band, \accIB{} is comparable or even somewhat larger than \accB{}, depending on $n$.  Hence the interband contribution at low pump fields plays in important role in the HG emission, especially around the temporal peak ($t=0$~ps) of the pump pulse.
In contrast, for the higher field case (Fig.~\ref{fig:bulkmc4Kacc}(b), $E_0=81\kVcm$), the hh contribution is significantly larger for all harmonics, as is the contribution from \accB{}, i.e., the HG emission is dominated by the \edsr{} of the hh after they have entered the band, similar to the $T=300$~K case (although the lower temperature here also affects the acoustic phonon scattering rates and equilibrium Fermi-Dirac distribution).

\begin{figure}[!t]
	\includegraphics[keepaspectratio,width=0.5\textwidth]
	{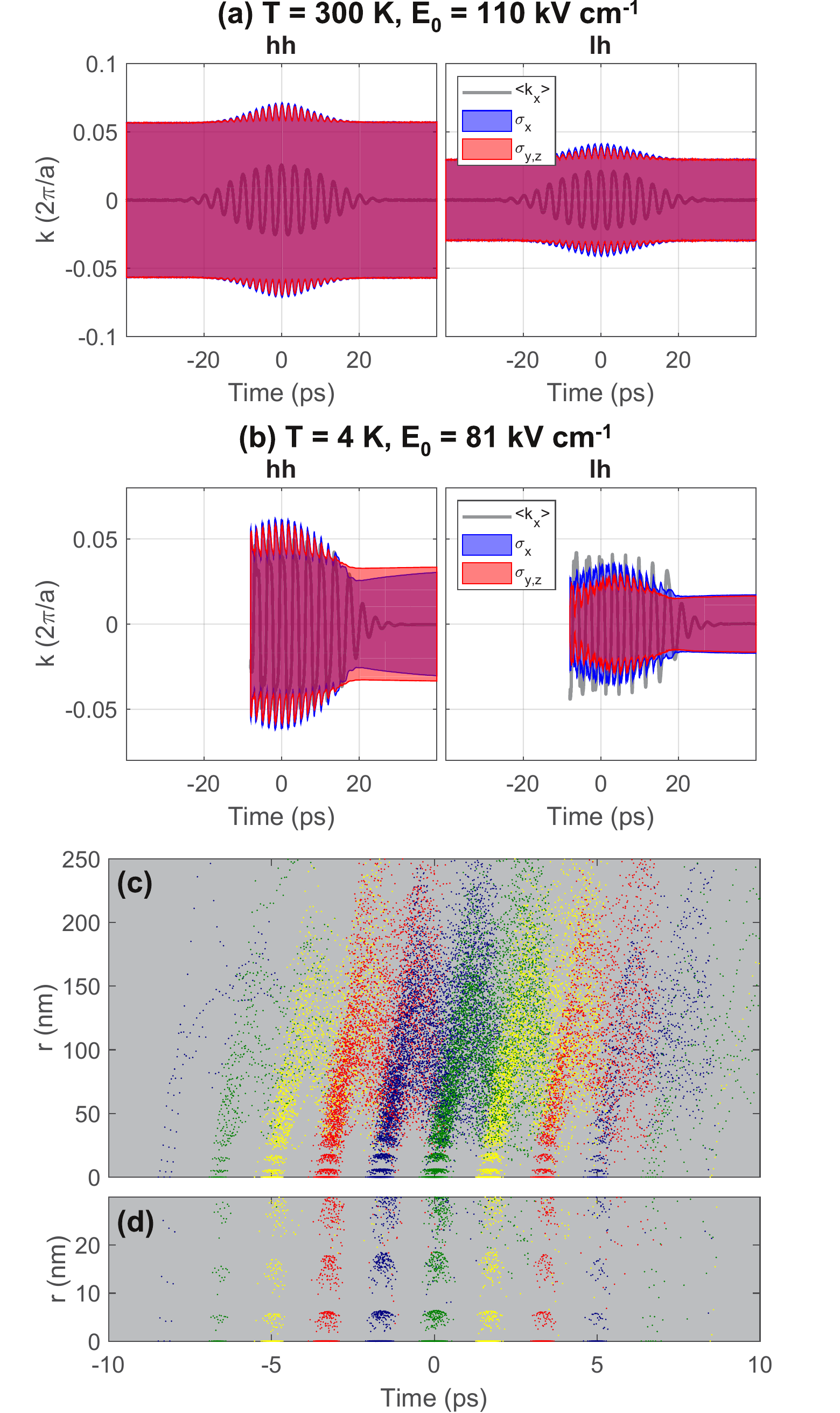}
	\caption{\label{fig:mctraj}
	Time-dependence of $k$-space distributions from local-MC simulations for (a) $T=300$~K and (b) $T=4$~K, for both hh (left panels) and lh (right panels), including ensemble average $k_x(t)$ (parallel to pump field), and rms spreads $\sigma_x(t)$ (along $k_x$) and $\sigma_y(t)=\sigma_z(t)$ (along $k_{y,z}$).  In (b), data truncated at early times before sufficient photoionized holes exist to perform statistics. (c) Time-dependent distance of hole from parent ion for the case in (b) (hh and lh combined), obtained by integrating $v_g(t)$ for each particle (608 holes in total, data only for the first 3~ps after respective ionization of each hole, downsampled to a time step of 50~fs for the first 500~fs, and 100~fs thereafter).  Color coding for each burst of photoionized holes during each pump half-cycle.  (d) Vertical zoom of (c) to allow inspection of small number of holes returning to parent ion during subsequent half-cycles 
 }
	\end{figure}

Finally, we address the evolution of the carrier distribution (both along $k_x$ parallel to the pump field, and transversely along $k_{y,z}$) and the real-space trajectories of photoionized holes, as predicted by the local-MC simulations. We consider only the high-field case.  
In Fig.~\ref{fig:mctraj}(a,b), we plot the time evolution of the ensemble-averaged wavevector $k_x(t)$ (separately for hh and lh), along with the rms spreads $\sigma_x(t)$ (along $k_x$) and $\sigma_{y,z}(t)$ (along $k_{y,z}$), for both the (a) room-temperature ($T=300$~K, Sec.~\ref{sec:hg300K}) and (b) low-temperature ($T=4$~K) situations.
For $T=300$~K, one sees again the relatively broad distribution before the pump pulse, due to the thermal Fermi-Dirac distribution, as discussed above in connection with Fig.~\ref{fig:bulkmc}(e,f), whose displacement/spread is only moderately perturbed during the pump excitation.  This is in contrast to the case we simulated in Ref.~\cite{Meng22} for $T=10$~K (assuming a small density of thermally ionized carriers) where $\sigma_{j}<0.01(2\pi/a)$ (for all $j=x,y,z$) before the pulse, and grew significantly during the pulse (although remaining somewhat smaller than the amplitude of $k_x$), as here the relative heating of the hole ensemble is small compared to $k_B T$. 
For the case of $T=4$~K in Fig.~\ref{fig:mctraj}(b), with holes generated by photoionization, one sees that the spread $\sigma_j$ rapidly acquires a comparable magnitude as for $T=300$~K, which is caused by rapid scattering processes even during the first half-cycle of the pump wave following respective ionization of each hole, with the spread $\sigma_{y,z}(t)$ pursuing $\sigma_{x}(t)$ closely due to transverse scattering events.
Although we defer the discussion of subsequent recollision of ionized holes with their parent ions to the next section (see Sec.~\ref{recoll}), we show here in Fig.~\ref{fig:mctraj}(c) results for the distance 
$r=|\vec{r}_h - \vec{r}_{\mathrm{B}^-}|$ between holes and their parent ions following ionization (608 holes tracked in total).
One can see clearly the coherent bursts of new holes about the field-peak of each pump half-cycle, and the ballistic initial acceleration to distances $r\sim 50$~nm, which then becomes diffuse due to momentum scattering, although one can still perceive a wave-like trend in each half-cycle as the holes accelerate back in the direction of their parent ions. 
To inspect for possible recollision, in Fig.~\ref{fig:mctraj}(d) we plot a vertical zoom of the data in (c).  
A close inspection (in particular for the half-cycles in the range of a few ps about $t=0$ where one has $\sim$100 ionized holes/burst) shows that a small fraction do return to distances $r<30$~nm, although very few $r<10$~nm, as the transverse momentum acquired from scattering causes them to pass $x=0$ displaced from $y=z=0$ (the same applies for holes returning after two half-cycles from the opposite $x$-direction).
Hence scattering is seen to seriously degrade the probability of a close recollision for the ionized holes.
The fact that the first near-recollisions occur close to a \textit{half}-cycle after ionization might at first seem unexpected, as it is well known from the semi-classical treatment of gas-phase recollision \cite{Lev94} that ionized charges created close to the field peaks, following the ballistic equation $\partial_t v = q E(t)/m$, return after a \textit{full} cycle.  However, one can show for the case with scattering, e.g. taking a simple constant damping rate $\Gamma$ ($\partial_t v = q E(t)/m - \Gamma v$), that the holes indeed re-approach their parent ions after a half-cycle for $\Gamma\ll \omega_1$ (see Supplementary).

\section{Discussion}\label{sec:discussion}
\subsection{Comparison with harmonic generation in graphene}
We first address the magnitude of the band nonlinearities determined for the experiments at $T=300$~K (Sec.~\ref{sec:hg300K}, Fig.~\ref{fig:pumpdep}(c)).
Here it is instructive to compare these values of \chin{n} with those determined recently for graphene \cite{Hafe18, Hafe19}, where the HG mechanism was attributed to an intra-cycle instability in the Drude heating/absorption of the carrier distribution.
Taking the effective thickness of the monolayer as $L_g=0.3$~nm, the authors deduced values \chin{n}
(in respective SI units, \mVx{(n-1)}) of $1.7\cdot10^{-9}$ ($n=3$), $1.2\cdot10^{-22}$ ($n=5$) and $1.7\cdot10^{-38}$ ($n=7$).
While the value of \chin{3} is 4 orders of magnitude higher than our present case, this is essentially due to the fact that graphene is an extremely potent electronic system, where the 2D carrier density in \cite{Hafe18} (due to substrate-induced electrostatic p-doping) of $N_{2D}=2.1\cdot10^{12}~\pcms$ is concentrated in a single monolayer.  If we normalize the nonlinear coefficients to the 3D carrier densities $N_d$ (Si:B) and $N_g=N_{2D}/L_g$ (graphene), one arrives at 
$\chin{3}/N_d=1.7\cdot10^{-35}\mVctn$ and $\chin{3}/N_g=2.4\cdot10^{-35}\mVctn$, i.e., the nonlinear response per carrier is almost equal.
One also notes that the ratios between successive $\chin{n}$ values ($n=3:5:7$) are also loosely correlated, in SI units $\sim1:10^{-14}:5\cdot10^{-14}$ for Si:B, and $\sim1:7\cdot10^{-14}:10^{-16}$ for graphene.
It is hence not correct to conclude that the unique band structure of graphene would lead directly or indirectly to a stronger nonlinear response per charge carrier as compared to conventional semiconductors.
While beyond the scope of the current work, this comparison raises the interesting question whether a more universal sum-rule \cite{Kuzyk13,Passos21} may apply for the (odd-order) THz nonlinear susceptibilities of charges in solid-state bands.

\subsection{\label{recoll}Contribution of recollisions to the nonlinear response?}
We turn now to the low-temperature HG process with photoionization. 
As mentioned in Sec.~\ref{sec:hg4K}, one can compare our scenario with that of the more well-established HG process observed for electron-hole pairs in solids with an energy bandgap, following interband excitation via either tunnel ionization \cite{Ghimire2011,vampa14,vampa15,You2017} or optical pre-excitation \cite{banks13,Wang2017}.
These studies provide strong support for recollision as a dominant mechanism for HG emission (in that context referred to as ``inter-band" generation), despite the presence of scattering processes, although the role of the subsequent anharmonic ``intra-band" currents are also proposed to dominate, at least in certain regimes \cite{Kemper13}.

In the case with pump fields at mid-infrared or higher frequencies (and hence carrier periods $<50$~fs), it seems reasonable that such a coherent recollision can occur for a significant fraction of the e-h pairs.  Nevertheless, even for recollision of e-h pairs (arising from tunnel ionization of optically excited excitons) in GaAs/AlGaAs quantum wells \cite{banks13} with a 570-GHz driving field, the authors deduce that LO phonon emission there does not destroy the recollision process as the required kinetic energy ($\Eopph{}=36$~meV) is only acquired during a short time directly before recollision.  
Note that this is based on the assumption that the dominant trajectory for a given emitted photon energy $h\nu_{em}$ corresponds to \textit{complete} \mbox{e-h} recombination, i.e., $h\nu_{em}=I_p+\mathcal{E}_{r}$, where the kinetic energy upon recollision $\mathcal{E}_{r}$ is governed by their birth time in the driving field.
This assumption is deduced from the condition for coherent emission in the purely ballistic case, as in dilute gases \cite{Lev94}, and may well need revision when stochastic scattering processes occur such that the trajectories are no longer deterministic.

Our photoionization+MC results (Sec.~\ref{sec:hg4K}, Fig.~\ref{fig:mctraj}(c)) suggest that the scattering processes occurring during the hole trajectories severely disrupt the return paths to their parent ions (both in terms of proximity and coherence).  
A simple treatment of the scattering (see Supplementary) also indicates a decisive influence of scattering, as it results in a kinetic energy $\mathcal{E}_r$ during any residual recollision events lower than $0.1\cdot U_p$ which is compared to $3.17\cdot U_p$ for the ballistic case ($U_p$ being again the ponderomotive energy, see 2nd paragraph of Sec.~\ref{sec:hg4K}). We note in passing, that this should not result in a high-frequency cut-off of HG in the frequency range covered in this study ($4\thz$), as that cut-off in the emission photon energy is given by $I_p + \mathcal{E}_r$, where the ionization energy $I_p=45\meV\hateq 10.9\thz$ is already above the covered photon energy range.

Nevertheless, given that the MC simulations here underestimate the HG emission strength (Fig.~\ref{fig:pumpdep4K}(a)), further theoretical studies should aim at quantifying any residual HG from recollision.
More generally, the results here strongly motivate future efforts to refine the description of the photoionization and any interactions with the parent ions.  
While we have employed the literature static-field ionization rate and initially populated only the lh band \cite{Darg95,Nie84}, this approach should be compared to time-dependent quantum-mechanical treatments, e.g. propagating the time-dependent density-matrix (or semiconductor-Bloch) equations \cite{kita15,luu16,Wang2017,Yue20}.
Such simulations may require a detailed treatment of both (i) the excited bound acceptor states \cite{Buczko92} if these are involved as intermediate states during ionization \cite{kita15,sereb16}, and (ii) the subsequent scattering processes occurring in the bands (where our MC treatment here does include explicitly effects such as the angular dependence and Pauli blocking). \\

\section{Conclusion}

In conclusion, we have presented a combined experimental and theoretical study of harmonic generation in Si:B with multi-cycle high-field THz pump pulses, for both temperature regimes where the band holes are either initially thermally ionized ($T=300$~K), or photoionized during the pulse ($T=4$~K). 
Pumping at 300~GHz, we observed up to the 9th harmonic order at room temperature and up to the 13th order at cryogenic temperature.  
Near quantitative agreement with experiment was achieved with Monte-Carlo simulations, but only when one includes both lh and hh bands and treats the propagation effects rigorously.
This agreement allowed us elucidate the microscopic harmonic generation processes on the basis of simulations of the local dynamics, which showed that the nonlinear response of pre-existing holes are dominated by energy-dependent scattering (and not band non-parabolicity) in this excitation regime, whereas for photoionized holes, the initial inter-band scattering processes also play an important role, especially close to the ionization threshold.
We find that scattering during the first sub-cycle should strongly influence the trajectories of photoionized holes (and any coherent recollision with their parent ions), which strongly motivates further studies of the photoionization process and subsequent interactions with the parent ions for the case of harmonic generation in solids.
Here, THz pump THz-probe (and/or photocurrent-probe) experiments would also be invaluable.

The nonlinear susceptibility per charge carrier was shown to be comparable to that for harmonic generation in graphene. The latter's Dirac-type band structure does apparently not lead to a higher per-carrier nonlinear response than that found in our doped Si. For practical applications, if one creates a (micrometer) thin layer of dopants on Si surface by using ion implantation, one would expect a high harmonic generation from this doped layer, which might open a new platform for future Si-based free-space or on-chip frequency multipliers or frequency-comb generators. In addition, if one further employs the waveguides or resonators, one might even generate high harmonics with an electric pump field of less than $10\kVcm$.

We finally note that this work serves to validate the application of MC simulations at this level of description for even higher frequencies and field strengths than it was previously employed, which will become increasingly relevant for the description of future (opto-)electronic devices.

\section*{Acknowledgments}
We acknowledge funding by the German Research Foundation (DFG) under the contract RO 770/41 and via the Collaborative Research Center TRR 288 (422213477, project B08). Parts of this research were carried out at the ELBE Center for High-Power Radiation Sources Sources at the Helmholtz-Zentrum Dresden - Rossendorf e. V., a member of the Helmholtz Association; for more information about the facility, see DOI: 10.17815/jlsrf-2-58. The authors thank Alexej Pashkin for his help for the experiments. We thank the ELBE team for the operation of the TELBE facility.



\bibliography{TELBE_HHG}

\end{document}